\documentstyle[12pt]{article}

\begin{document}


\begin{center}
Quantum fluctuations of systems 
of interacting electrons in two spatial dimensions.
\end{center}
\begin{center}
Maciej M. Duras
\end{center}
\begin{center}
Institute of Physics, Cracow University of Technology, 
ulica Podchor\c{a}\.zych 1, PL-30084 Cracow, Poland.
\end{center}

\begin{center}
Email: mduras @ riad.usk.pk.edu.pl
\end{center}

\begin{center}
``International Workshop on
Critical Stability of Few-Body Quantum Systems'';
October 17, 2005 -  October 22, 2005;
Max Planck Institute for the Physics of Complex Systems,
Dresden, Germany (2005).
\end{center}

\begin{center}
AD 2005 October 5th
\end{center}

\section{Abstract}
\label{sec-Abstract}
The random matrix ensembles (RME) of quantum statistical Hamiltonian operators,
e.g. Gaussian random matrix ensembles (GRME) 
and Ginibre random matrix ensembles (Ginibre RME), 
are applied to following quantum statistical systems: 
nuclear systems, molecular systems, 
and two-dimensional electron systems (Wigner-Dyson electrostatic analogy). 
Measures of quantum chaos and quantum integrability 
with respect to eigenergies of quantum systems are defined and calculated. 
Quantum statistical information functional is defined as negentropy 
(opposite of entropy or minus entropy). 
The distribution function for the random matrix ensembles is derived 
from the maximum entropy principle. 

\section{Introduction}
\label{sec-introduction}

Random Matrix Theory (RMT) studies
as an example random matrix variables ${\cal H}$ corresponding 
to random quantum Hamiltonian operators $\hat {\cal{H}}$.
Their matrix elements
${\cal H}_{ij}, i, j =1,...,N, N \geq 1,$ are independent random scalar variables
and are associated to matrix elements $H_{ij}$ of Hamiltonian operator $\hat{H}$.
\cite{Haake 1990,Guhr 1998,Mehta 1990 0,Reichl 1992,Bohigas 1991,Porter 1965,Brody 1981,Beenakker 1997}.
We will use the following notation:
$\hat{H}$ is hermitean Hamiltonian operator,
$H$ is its hermitean matrix representation,
$E_{i}$'s are the eigenvalues of both $\hat{H}$ and $H$.
The matrix elements of $H$ are $H_{ij}, i, j =1,...,N, N \geq 1$.
Moreover, $\hat {\cal{H}}$ is random hermitean Hamiltonian operator,
${\cal H}$ is its hermitean random matrix representation.
The values assumed by $\hat {\cal{H}}$ are denoted by $\hat{H}$,
whereas the values of ${\cal{H}}$ are $H$, respectively.
The ${\cal E}_{i}$'s are the random eigenvalues of both $\hat {\cal{H}}$
and ${\cal{H}}$. The random matrix elements of ${\cal H}$ are
${\cal H}_{ij}, i, j =1,...,N, N \geq 1$.
Both ${\cal E}_{i}$ and ${\cal H}_{ij}$ are random variables,
whereas ${\cal H}$ is random matrix variable.
The values assumed by random variables ${\cal E}_{i}$
are denoted $E_{i}$.
There were studied among others the following
Gaussian Random Matrix ensembles GRME:
orthogonal GOE, unitary GUE, symplectic GSE,
as well as circular ensembles: orthogonal COE,
unitary CUE, and symplectic CSE.
The choice of ensemble is based on quantum symmetries
ascribed to the Hamiltonian operator $\hat{H}$. 
The Hamiltonian operator $\hat{H}$
acts on quantum space $V$ of eigenfunctions.
It is assumed that $V$ is $N$-dimensional Hilbert space
$V={\bf F}^{N}$, where the real, complex, or quaternion
field ${\bf F}={\bf R, C, H}$,
corresponds to GOE, GUE, or GSE, respectively.
If the Hamiltonian matrix $H$ is hermitean $H=H^{\dag}$,
then the probability density function $f_{{\cal H}}$ of ${\cal H}$ reads:
\begin{eqnarray}
& & f_{{\cal H}}(H)={\cal C}_{{\cal H} \beta} 
\exp{[-\beta \cdot \frac{1}{2} \cdot {\rm Tr} (H^{2})]},
\label{pdf-GOE-GUE-GSE} \\
& & {\cal C}_{{\cal H} \beta}=(\frac{\beta}{2 \pi})^{{\cal N}_{{\cal H} \beta}/2}, 
\nonumber \\
& & {\cal N}_{{\cal H} \beta}=N+\frac{1}{2}N(N-1)\beta, \nonumber \\
& & \int f_{{\cal H}}(H) dH=1,
\nonumber \\
& & dH=\prod_{i=1}^{N} \prod_{j \geq i}^{N} 
\prod_{\gamma=0}^{D-1} dH_{ij}^{(\gamma)}, \nonumber \\
& & H_{ij}=(H_{ij}^{(0)}, ..., H_{ij}^{(D-1)}) \in {\bf F}, \nonumber
\end{eqnarray}
where the parameter $\beta$ assume values 
$\beta=1,2,4,$  for GOE($N$), GUE($N$), GSE($N$), respectively,
and ${\cal N}_{{\cal H} \beta}$ is number of independent matrix elements $H_{ij}$
of hermitean Hamiltonian matrix $H$.
The Hamiltonian $H$ belongs to Lie group of hermitean $N \times N $ dimensional ${\bf F}$-matrices,
and the matrix Haar's measure $dH$ is invariant under
transformations from the unitary group U($N$, {\bf F}).
The eigenenergies $E_{i}, i=1, ..., N$, of $H$, are real-valued
random variables $E_{i}=E_{i}^{\star}$, and also for the
random eigenenergies it holds: ${\cal E}_{i}={\cal E}_{i}^{\star}$.
It was Eugene Wigner who firstly dealt with eigenenergy level repulsion
phenomenon studying nuclear spectra \cite{Haake 1990,Guhr 1998,Mehta 1990 0}.
RMT is applicable now in many branches of physics:
nuclear physics (slow neutron resonances, highly excited complex nuclei),
condensed phase physics (fine metallic particles,  
random Ising model [spin glasses]),
quantum chaos (quantum billiards, quantum dots), 
disordered mesoscopic systems (transport phenomena),
quantum chromodynamics, two-dimensional Euclidean quantum gravity (2D EQG), 
Euclidean field theory (EFT).

\section{The Ginibre ensembles}
\label{sec-ginibre-ensembles}

Jean Ginibre considered another example of GRME
dropping the assumption of hermiticity of Hamiltonians
thus defining generic ${\bf F}$-valued Hamiltonian matrix $K$
\cite{Haake 1990,Guhr 1998,Ginibre 1965,Mehta 1990 1}.
Hence, $K$ belong to general linear Lie group GL($N$, {\bf F}),
and the matrix Haar's measure $dK$ is invariant under
transformations from that group.
The distribution $f_{{\cal K}}$ of ${\cal K}$ is given by:
\begin{eqnarray}
& & f_{{\cal K}}(K)={\cal C}_{{\cal K} \beta} 
\exp{[-\beta \cdot \frac{1}{2} \cdot {\rm Tr} (K^{\dag}K)]},
\label{pdf-Ginibre} \\
& & {\cal C}_{{\cal K} \beta}=(\frac{\beta}{2 \pi})^{{\cal N}_{{\cal K} \beta}/2}, 
\nonumber \\
& & {\cal N}_{{\cal K} \beta}=N^{2}\beta, \nonumber \\
& & \int f_{{\cal K}}(K) dK=1,
\nonumber \\
& & dK=\prod_{i=1}^{N} \prod_{j=1}^{N} 
\prod_{\gamma=0}^{D-1} dK_{ij}^{(\gamma)}, \nonumber \\
& & K_{ij}=(K_{ij}^{(0)}, ..., K_{ij}^{(D-1)}) \in {\bf F}, \nonumber
\end{eqnarray}
where $\beta=1,2,4$, stands for real, complex, and quaternion
Ginibre ensembles, respectively,
and ${\cal K}$ is random matrix variable associated with matrix $K$.
Therefore, the eigenenergies ${\cal Z}_{i}$ of quantum system 
ascribed to Ginibre ensemble are complex-valued random variables.
The eigenenergies ${\cal Z}_{i}, i=1, ..., N$,
of nonhermitean random matrix Hamiltonian ${\cal K}$ are not real-valued random variables
${\cal Z}_{i} \neq {\cal Z}_{i}^{\star}$.
Jean Ginibre postulated the following
joint probability density function 
of random vector ${\cal Z}$ of complex eigenvalues ${\cal Z}_{1}, ..., {\cal Z}_{N}$
for $N \times N$ dimensional random matrix Hamiltonians ${\cal K}$ for $\beta=2$
\cite{Haake 1990,Guhr 1998,Ginibre 1965,Mehta 1990 1}:
\begin{eqnarray}
& & P(Z_{1}, ..., Z_{N})=
\label{Ginibre-joint-pdf-eigenvalues} \\
& & =\prod _{j=1}^{N} \frac{1}{\pi \cdot j!} \cdot
\prod _{i<j}^{N} \vert Z_{i} - Z_{j} \vert^{2} \cdot
\exp (- \sum _{j=1}^{N} \vert Z_{j}\vert^{2}),
\nonumber
\end{eqnarray}
where $Z_{i}$ are complex-valued sample points (values) of ${\cal Z}_{i}$
($Z_{i} \in {\bf C}$).
 
We emphasize here Wigner and Dyson's electrostatic analogy.
A Coulomb gas of $N$ unit charges $Q_{i}$ moving on complex plane (Gauss's plane)
{\bf C} is considered. The complex vectors of positions
of charges are $Z_{i}$ and potential energy $U$ of the system is:
\begin{equation}
U(Z_{1}, ...,Z_{N})=
- \sum_{i<j}^{N} \ln \vert Z_{i} - Z_{j} \vert
+ \frac{1}{2} \sum_{i} \vert Z_{i} \vert ^{2}. 
\label{Coulomb-potential-energy}
\end{equation}
If gas is in thermodynamical equilibrium at temperature
$T= \frac{1}{2 k_{B}}$ 
($\beta= \frac{1}{k_{B}T}=2$, $k_{B}$ is Boltzmann's constant),
then probability density function of vectors $Z_{i}$ of positions is 
$P(Z_{1}, ..., Z_{N})$ Eq. (\ref{Ginibre-joint-pdf-eigenvalues}).
Therefore, complex eigenenergies $Z_{i}$ of quantum system 
are analogous to vectors of positions of charges of Coulomb gas.
Moreover, complex-valued spacings $\Delta^{1} Z_{i}$ 
(first order forward/progressive finite differences)
of complex eigenenergies $Z_{i}$ of quantum system:
\begin{equation}
\Delta^{1} Z_{i}=\Delta Z_{i}=Z_{i+1}-Z_{i}, i=1, ..., (N-1),
\label{first-diff-def}
\end{equation}
are analogous to vectors of relative positions of electric charges.
Finally, complex-valued
second differences $\Delta^{2} Z_{i}$ 
(second order forward/progressive finite differences)
of complex eigenenergies $Z_{i}$:
\begin{equation}
\Delta ^{2} Z_{i}=Z_{i+2} - 2Z_{i+1} + Z_{i}, i=1, ..., (N-2),
\label{Ginibre-second-difference-def}
\end{equation}
are analogous to
vectors of relative positions of vectors
of relative positions of electric charges.

The eigenenergies $Z_{i}=Z(i)$ can be treated as values of function $Z$
of discrete parameter $i=1, ..., N$.
The ``Jacobian'' of $Z_{i}$ reads:
\begin{equation}
{\rm Jac} Z(i)= {\rm Jac} Z_{i}= \frac{\partial Z_{i}}{\partial i}
= \frac{d Z_{i}}{d i}
\simeq \frac{\Delta^{1} Z_{i}}{\Delta^{1} i}=
\frac{\Delta Z_{i}}{\Delta i}=\Delta^{1} Z_{i},
\label{jacobian-Z}
\end{equation}
where $\Delta i=i+1-1=1$.
We readily have, that the spacing is an discrete analog of Jacobian,
since the indexing parameter $i$ belongs to discrete space
of indices $i \in I=\{1, ..., N \}$. Therefore, the first derivative 
$\frac{\Delta Z_{i}}{\Delta i}$
with respect to $i$ reduces to the first forward (progressive) differential quotient
$\frac{\Delta Z_{i}}{\Delta i}$.
The Hessian is a Jacobian applied to Jacobian.
We immediately have the formula for discrete ``Hessian'' for the eigenenergies $Z_{i}$:
\begin{equation}
{\rm Hess} Z(i)={\rm Hess} Z_{i}= \frac{\partial ^{2} Z_{i}}{\partial i^{2}}
= \frac{d ^{2} Z_{i}}{d i^{2}}
\simeq \frac{\Delta^{2} Z_{i}}{\Delta^{1} i^{2}}=
\frac{\Delta^{2} Z_{i}}{(\Delta i)^{2}} = \Delta^{2} Z_{i}.
\label{hessian-Z}
\end{equation}
Thus, the second difference of $Z$ is discrete analog of Hessian of $Z$.
One emphasizes that both ``Jacobian'' and ``Hessian''
work on discrete index space $I$ of indices $i$.
The spacing is also a discrete analog of energy slope
whereas the second difference corresponds to
energy curvature with respect to external parameter $\lambda$
describing parametric ``evolution'' of energy levels
\cite{Zakrzewski 1,Zakrzewski 2}.
The finite differences of order higher than two
are discrete analogs of compositions of ``Jacobians'' with ``Hessians'' of $Z$.

The eigenenergies $E_{i}, i \in I$, of the hermitean Hamiltonian matrix $H$
are ordered increasingly real numbers.
They are values of discrete function $E_{i}=E(i)$.
The first order progressive differences of adjacent eigenenergies:
\begin{equation}
\Delta^{1} E_{i}=E_{i+1}-E_{i}, i=1, ..., (N-1),
\label{first-diff-def-GRME}
\end{equation}
are analogous to vectors of relative positions of electric charges
of one-dimensional Coulomb gas. It is simply the spacing of two adjacent
energies.
Real-valued
progressive finite second differences $\Delta^{2} E_{i}$ of eigenenergies:
\begin{equation}
\Delta ^{2} E_{i}=E_{i+2} - 2E_{i+1} + E_{i}, i=1, ..., (N-2),
\label{Ginibre-second-difference-def-GRME}
\end{equation}
are analogous to vectors of relative positions 
of vectors of relative positions of charges of one-dimensional
Coulomb gas.
The $\Delta ^{2} Z_{i}$ have their real parts
${\rm Re} \Delta ^{2} Z_{i}$,
and imaginary parts
${\rm Im} \Delta ^{2} Z_{i}$, 
as well as radii (moduli)
$\vert \Delta ^{2} Z_{i} \vert$,
and main arguments (angles) ${\rm Arg} \Delta ^{2} Z_{i}$.
$\Delta ^{2} Z_{i}$ are extensions of real-valued second differences:
\begin{equation}
{\rm Re} (\Delta ^{2} Z_{i})={\rm Re}(Z_{i+2}-2Z_{i+1}+Z_{i})
=\Delta ^2 {\rm Re} Z_{i}= \Delta ^{2} E_{i}, i=1, ..., (N-2),
\label{second-diff-def}
\end{equation}
of adjacent ordered increasingly real-valued eigenenergies $E_{i}$
of Hamiltonian matrix $H$ defined for
GOE, GUE, GSE, and Poisson ensemble PE
(where Poisson ensemble is composed of uncorrelated
randomly distributed eigenenergies)
\cite{Duras 1996 PRE,Duras 1996 thesis,Duras 1999 Phys,Duras 1999 Nap,Duras 2000 JOptB,Duras 2001 Vaxjo,Duras 2001
Pamplona, Duras 2003 Spie03, Duras 2004 Spie04, Duras 2005 Spie05}.
The ``Jacobian'' and ``Hessian'' operators of energy function $E(i)=E_{i}$
for these ensembles read:
\begin{equation}
{\rm Jac} E(i)= {\rm Jac} E_{i}= \frac{\partial E_{i}}{\partial i}
= \frac{d E_{i}}{d i}
\simeq \frac{\Delta^{1} E_{i}}{\Delta^{1} i}=
\frac{\Delta E_{i}}{\Delta i}=\Delta^{1} E_{i},
\label{jacobian-E}
\end{equation}
and
\begin{equation}
{\rm Hess} E(i)={\rm Hess} E_{i}= \frac{\partial ^{2} E_{i}}{\partial i^{2}}
= \frac{d ^{2} E_{i}}{d i^{2}}
\simeq \frac{\Delta^{2} E_{i}}{\Delta^{1} i^{2}}=
\frac{\Delta^{2} E_{i}}{(\Delta i)^{2}} = \Delta^{2} E_{i}.
\label{hessian-E}
\end{equation}
The treatment of first and second differences of eigenenergies
as discrete analogs of Jacobians and Hessians
allows one to consider these eigenenergies as a magnitudes 
with statistical properties studied in discrete space of indices.
The labelling index $i$ of the eigenenergies is
an additional variable of ``motion'', hence the space of indices $I$
augments the space of dynamics of random magnitudes.

One may also study the finite expressions of random eigenenergies
and their distributions. The finite expressions are more general than finite
difference quotients and they represent the derivatives of eigenenergies
with respect to labelling index $i$ more accurately
\cite{Collatz 1955, Collatz 1960}.  
 
\section{The Maximum Entropy Principle}
\label{sec-maximal-entropy}
In order to derive the probability distribution
in matrix space 
${\cal M}={\rm MATRIX}(N, N, {\bf F})$ of all $N \times N$ dimensional
${\bf F}$-valued matrices $X$
we apply the maximum entropy principle:
\begin{equation}
{\rm max} \{S_{\beta}[f_{{\cal X}}]: \left< 1 \right>=1, 
\left< h_{{\cal X}} \right>=U_{\beta} \},
\label{maximal-entropy-problem}
\end{equation}
whereas $\left< ... \right>$ denotes the random matrix ensemble average,
\begin{equation}
\left< g_{{\cal X}} \right>
=\int_{{\cal M}} g_{{\cal X}}(X) f_{{\cal X}}(X) d X,
\label{g-average-definition}
\end{equation}
and 
\begin{equation}
S_{\beta}[f_{{\cal X}}]=\left< - k_{B} \ln f_{{\cal X}} \right>
= \int_{{\cal M}} (- k_{B} \ln f_{{\cal X}}(X)) f_{{\cal X}}(X) d X,
\label{entropy-functional-definition}
\end{equation}
is the entropy functional,
\begin{equation}
\left< h_{{\cal X}} \right>
= \int_{{\cal M}} h_{{\cal X}}(X) f_{{\cal X}}(X) d X,
\label{h-average-definition}
\end{equation}
and 
\begin{equation}
\left< 1\right>
= \int_{{\cal M}} 1 f_{{\cal X}}(X) d X =1.
\label{f-normalization-definition}
\end{equation}
Here, $h_{{\cal X}}$ stands for ``microscopic potential energy''
of random matrix variable ${\cal X}$, and 
$U_{\beta}$ is ``macroscopic potential energy''.
We recover the Gaussian random matrix ensemble distribution 
$f_{{\cal H}}$ Eq. (\ref{pdf-GOE-GUE-GSE}),
or Ginibre random matrix ensemble distribution
$f_{{\cal K}}$ Eq. (\ref{pdf-Ginibre}), respectively,
if we put 
$h_{{\cal X}}(X)= {\rm Tr} (\frac{1}{2} X^{\dag}X),$
and $X=H,$ or $X=K$, respectively.
The maximization of entropy 
$S_{\beta}[f_{{\cal X}}]$ 
under two additional
constraints of normalization of the probability density function 
$f_{{\cal X}}$ of ${\cal X}$,
and of equality of its first momentum and ``macroscopic potential energy'',
is equivalent to the minimization of the following functional
${\cal F}[f_{{\cal X}}]$ with the use of Lagrange multipliers 
$\alpha_{1}, \beta_{1}$: 
\begin{eqnarray}
& & {\rm min} \{ {\cal F} [f_{{\cal X}}] \},
\label{maximal-entropy-problem-Lagrange} \\
& & {\cal F} [f_{{\cal X}}] 
= \int_{{\cal M}} ( +k_{B} \ln f_{{\cal X}}(X)) f_{{\cal X}}(X) d X
+\alpha_{1} \int_{{\cal M}} f_{{\cal X}}(X) d X \nonumber \\
& & + \beta_{1} \int h_{{\cal X}}(X) f_{{\cal X}}(X) d X = \nonumber \\
& & = -S_{\beta}[f_{{\cal X}}] + \alpha_{1} \left< 1 \right> 
+ \beta_{1} \left< h_{{\cal X}} \right> . \nonumber
\end{eqnarray}
It follows, that the first variational derivative of ${\cal F}[f_{{\cal X}}]$
must vanish:
\begin{equation}
\frac{\delta {\cal F} [f_{{\cal X}}]}{\delta f_{{\cal X}}}=0,
\label{Lagrange-first-derivative}
\end{equation}
which produces:
\begin{equation}
k_{B} (\ln f_{{\cal X}}(X) + 1) 
+\alpha_{1} + \beta_{1} h_{{\cal X}}(X)=0,
\label{Lagrange-integrand}
\end{equation}
and equivalently:
\begin{eqnarray}
& & f_{{\cal X}}[X]={\cal C}_{{\cal X} \beta}  \cdot
\exp{[-\beta \cdot h_{{\cal X}}(X)]}
\label{pdf-GOE-GUE-GSE-PH-partition-function-Lagrange} \\
& & {\cal C}_{{\cal X} \beta}= \exp[ -(\alpha_{1}+1) \cdot k_{B}^{-1}],
\beta=\beta_{1} \cdot k_{B}^{-1}.
\nonumber
\end{eqnarray}
One easily shows that
\begin{equation}
\frac{\delta ^2 {\cal F} [f_{{\cal X}}]}{\delta f_{{\cal X}}^2} > 0.
\label{Lagrange-second-derivative}
\end{equation}
The variational principle of maximum entropy does not
force additional condition on functional form 
of $h_{{\cal X}}(X)$. 
The quantum statistical information functional $I_{\beta}$ 
is the opposite of entropy:
\begin{equation}
I_{\beta}[f_{{\cal X}}]=-S_{\beta}[f_{{\cal X}}]
= \int_{{\cal M}} ( + k_{B} \ln f_{{\cal X}}(X)) f_{{\cal X}}(X) d X.
\label{information-entropy}
\end{equation}
Information is negentropy, and entropy is neginformation.
The maximum entropy principle is equivalent to
minimum information principle.


\end{document}